\documentclass[11pt,twoside]{article}
\usepackage{asp2010}

\resetcounters

\begin{document}

\title{Magnetic field amplification by cosmic rays in supernova remnants}
\author{K.M.~Schure,$^1$ and A.R.~Bell,$^1$
\affil{$^1$Department of Physics, University of Oxford, Clarendon Laboratory, Parks Road, Oxford OX1 3PU, United Kingdom}}

\begin{abstract}
Magnetic field amplification is needed to accelerate cosmic cays to PeV energies in supernova remants. Escaping cosmic rays trigger a return current in the plasma that drives a non-resonant hybrid instability. 
We run simulations in which we represent the escaping cosmic rays with the plasma return current, keeping the maximum cosmic ray energy fixed, and evaluate its effects on the upstream medium. In addition to magnetic field amplification, density perturbations arise that, when passing through the shock, further increase amplification levels downstream.
As the growth rate of the instability is most rapid for the smaller scales, the resolution is a limiting factor in the amplification that can be reached with these simulations. 
\end{abstract}

\section{Introduction}
Cosmic rays (CRs) are accelerated in supernova remnants (SNRs). They can be observed in situ mainly in the form of relativistic electrons, that radiate synchrotron radiation from radio- to X-ray wavelengths \citep{1992ReynoldsEllison,1995Koyamaetal}. From the radiation signature the energies for electrons and magnetic field strengths can be deduced \citep{2003VinkLaming}. Acceleration of electrons to TeV energies seems ubiquitous, and diffusive shock acceleration is the preferred acceleration method. Protons are also believed to be accelerated in this process, possibly to higher energies than the electrons, as they do not suffer the same amount of losses. They could leave a signature in gamma rays, which is, however, difficult to disentangle from inverse compton emission of the electron population. Future observations of high-energy gamma rays by CTA \citep{2011CTA}, in combination with theory, could shed light on this ambiguity.

For efficient acceleration, a magnetic field that is strong and turbulent enough needs to be present upstream of the shock, such that the cosmic rays are scattered back and forth across the shock, gaining energy at each shock crossing. The most likely candidate to amplify the field is the non-resonant hybrid instability that grows most rapidly on small scales \citet{2004Bell}. We present simulations of this instability in the upstream of a SNR shock wave. It can be shown to be responsible for creating significant amplification upstream of the shock. In addition to amplified magnetic field it creates density inhomogeneities. When the shock crosses this region of density and magnetic field fluctuations, additional amplification occurs downstream \citep{2011Beresnyak,2012Guoetal}.

\section{Current driven MFA}

If cosmic rays are efficiently accelerated at the shock, it can be assumed that a certain fraction $\chi$ of the shock's kinetic energy is transferred to the relativistic particle population. If this fraction is independent on what happens at the high-energy end of the cosmic ray spectrum, the escaping cosmic rays will give rise to a current that decreases when the maximum particle energy increases. The return current that is consequently triggered in the plasma can give rise to a number of current driven instabilities \citep[for an overview see][]{2012Schureetal}.

We are investigating the effect of this current on the magnetic field growth and stability of the upstream plasma in the regime of the non-resonant hybrid instability \citep{2004Bell}. Its maximum growth rate and corresponding wave number are: \begin{eqnarray}
\gamma_{max}=\frac{j}{c}\sqrt{\frac{\pi}{\rho}}, \qquad k_{max}=\frac{2 \pi j}{B c}.
\end{eqnarray}
In order to be able to resolve this wavenumber in simulations, the size of the grid cells needs to be smaller than $\Delta x_g \le 2 \pi /k$.

\begin{figure*}[!tp]
  \centering
\includegraphics[width=0.32\textwidth]{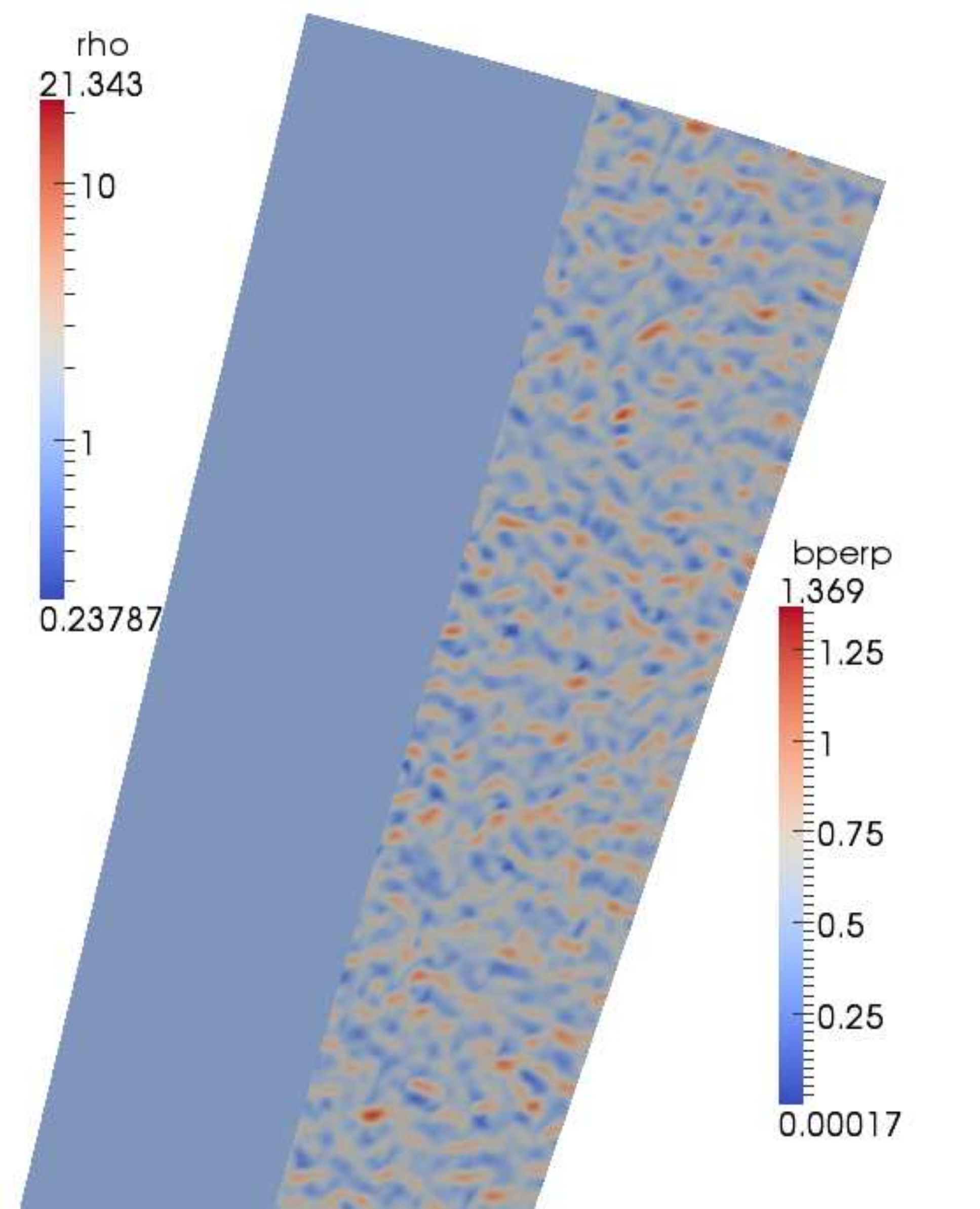}
\includegraphics[width=0.32\textwidth]{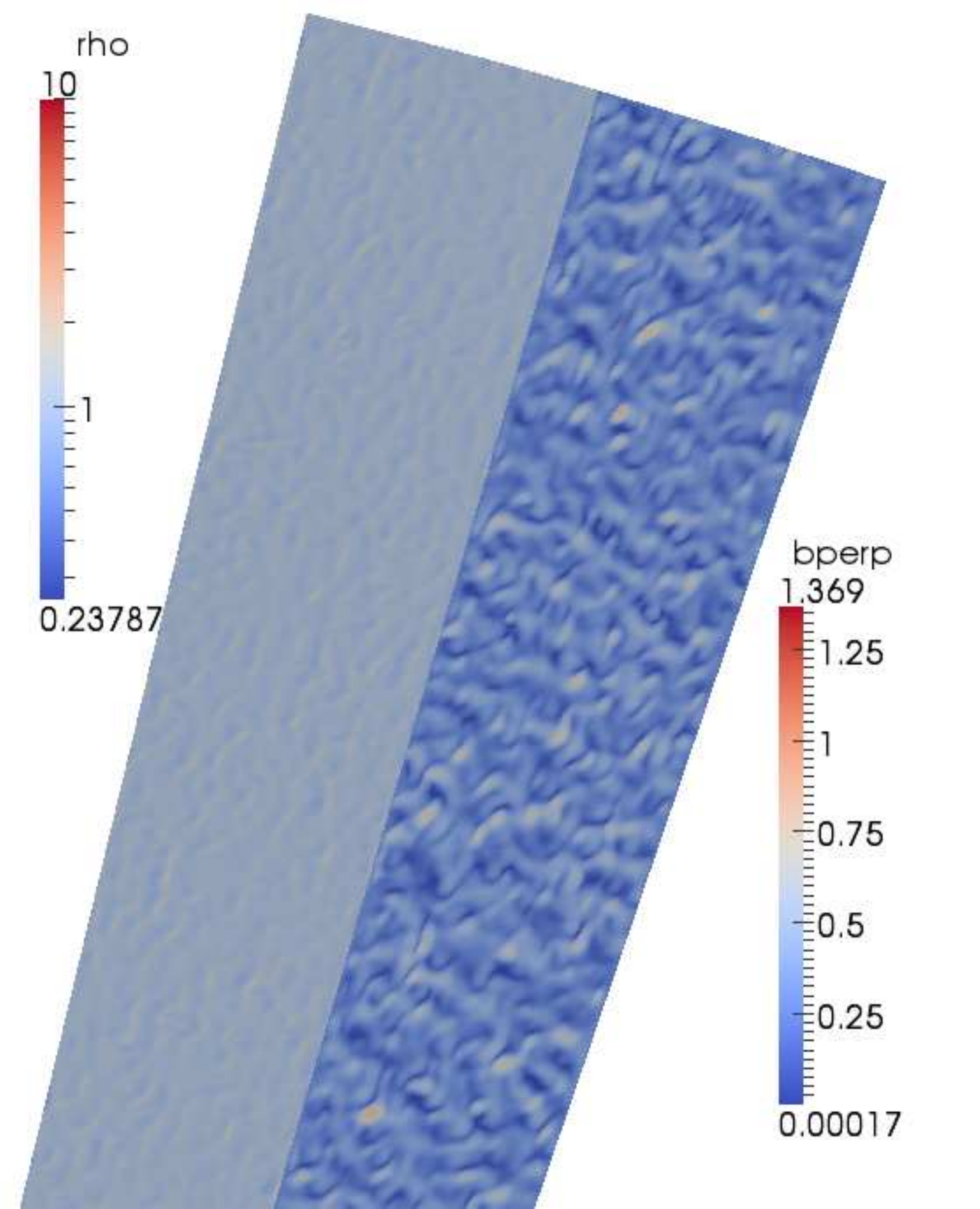}
\includegraphics[width=0.32\textwidth]{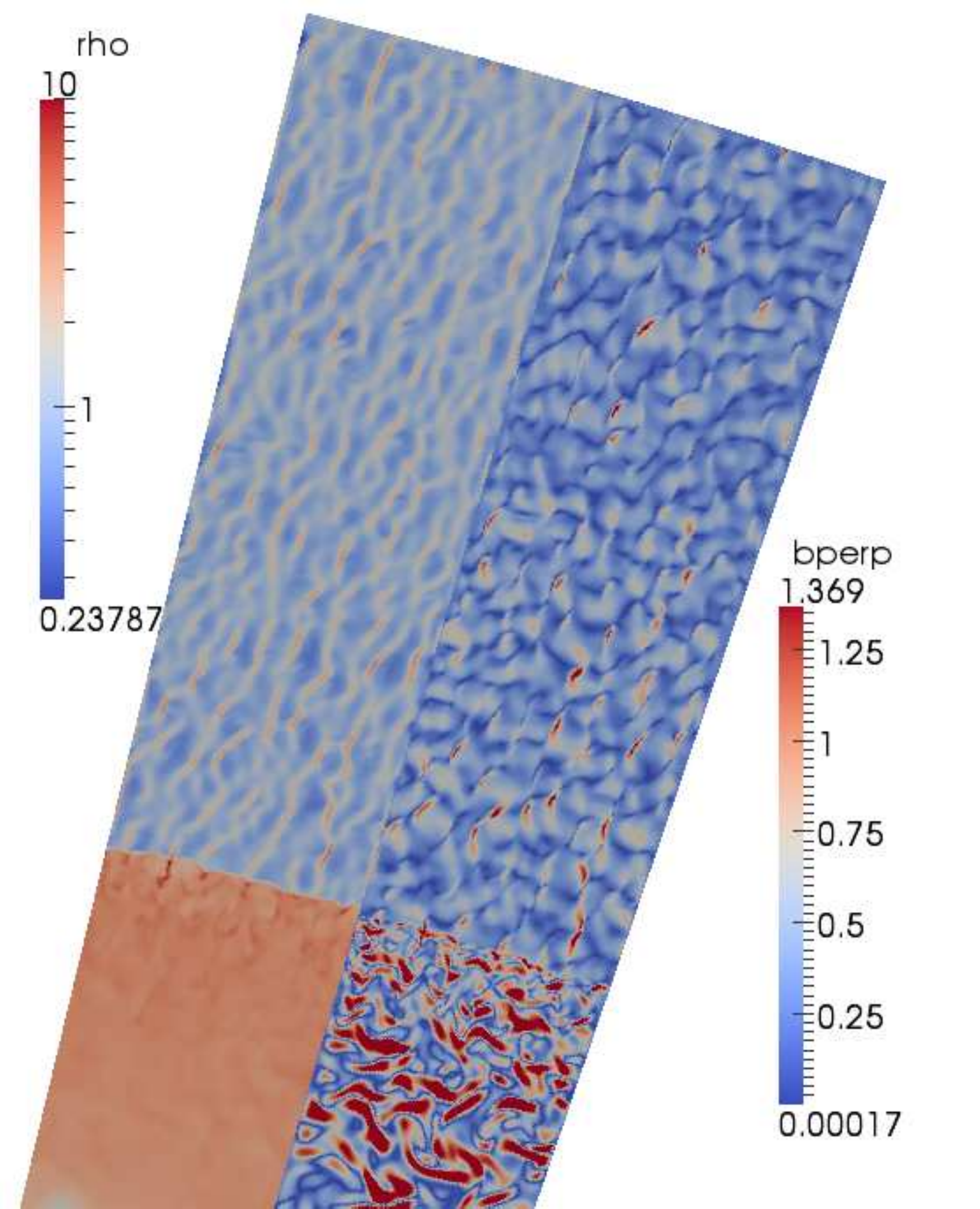}
 \caption[ ] {Evolution of the upstream medium of a SNR, cropped at the bottom. Both the density and $B_\perp$ are shown, time progresses left to right (0, 80, 160 yrs). The angle that is simulated is $\sim 3^\circ$ wide, corresponding to a physical length scale of $\sim 2.4 \times 10^{17}$~cm at the maximum radius that is plotted ($5 \times 10^{18}$~cm).
    \label{fig:largescale}}
  \end{figure*}

We have run a simulation in which we let a SNR evolve in a medium that initially has a homogeneous density, and a background magnetic field $B_0$, to which we add a turbulent magnetic field with fluctuations on rather small scales, such that they are resolved by a few grid cells. We use the MHD code from the AMRVAC framework \citep{2007HolstKeppens} on a 2.5D spherical grid, with symmetry in the 3rd dimension. $B_0$ is in the direction of the symmetry axis. Because of resolution requirements we only take a small wedge, over an angle $\sim3^\circ$ and radial extent of $5\times 10^{18}$~cm. In this case, the smallest grid size is approximately $5 \times 10^{14}$~cm, which is inversely linked to the maximum wavenumber of the instability that we can expect to grow. The current is representative for particles escaping beyond $10^{14}$~eV, with $30$\% of the shock's kinetic energy going into cosmic rays. The fastest growing modes is parallel to the zeroth-order magnetic field, and it drives growth of the magnetic field perpendicular to the zeroth order current ($B_\perp$). As the growth rate increases with wave number, the limited resolution in this simulation means that we do not expect to resolve amplification at the maximum level. The general trend though can be seen in the results presented in Figure~\ref{fig:largescale}, which shows the evolution of the density and $B_\perp$ with time. Only a section of the simulated region is shown, such that the shock can be seen to enter the frame only at the last shown time step. 

The Lorentz force resulting from the $j_0 \times B_\perp$ pushes around the plasma, thereby locally amplifying the magnetic field where the density is increased, while increasing the length scale of the structures. The maximum growth is limited because of the resolution and the small cosmic ray current: $\delta B/B_0$ grows only a factor of a few in the time available until the shock passes. When the growth of fluctuations is faster, the density perturbations can be seen to disturb the shock front.

\begin{figure*}[h]
  \centering
 \includegraphics[trim=30 80 0 80,clip=true,width=0.45\textwidth]{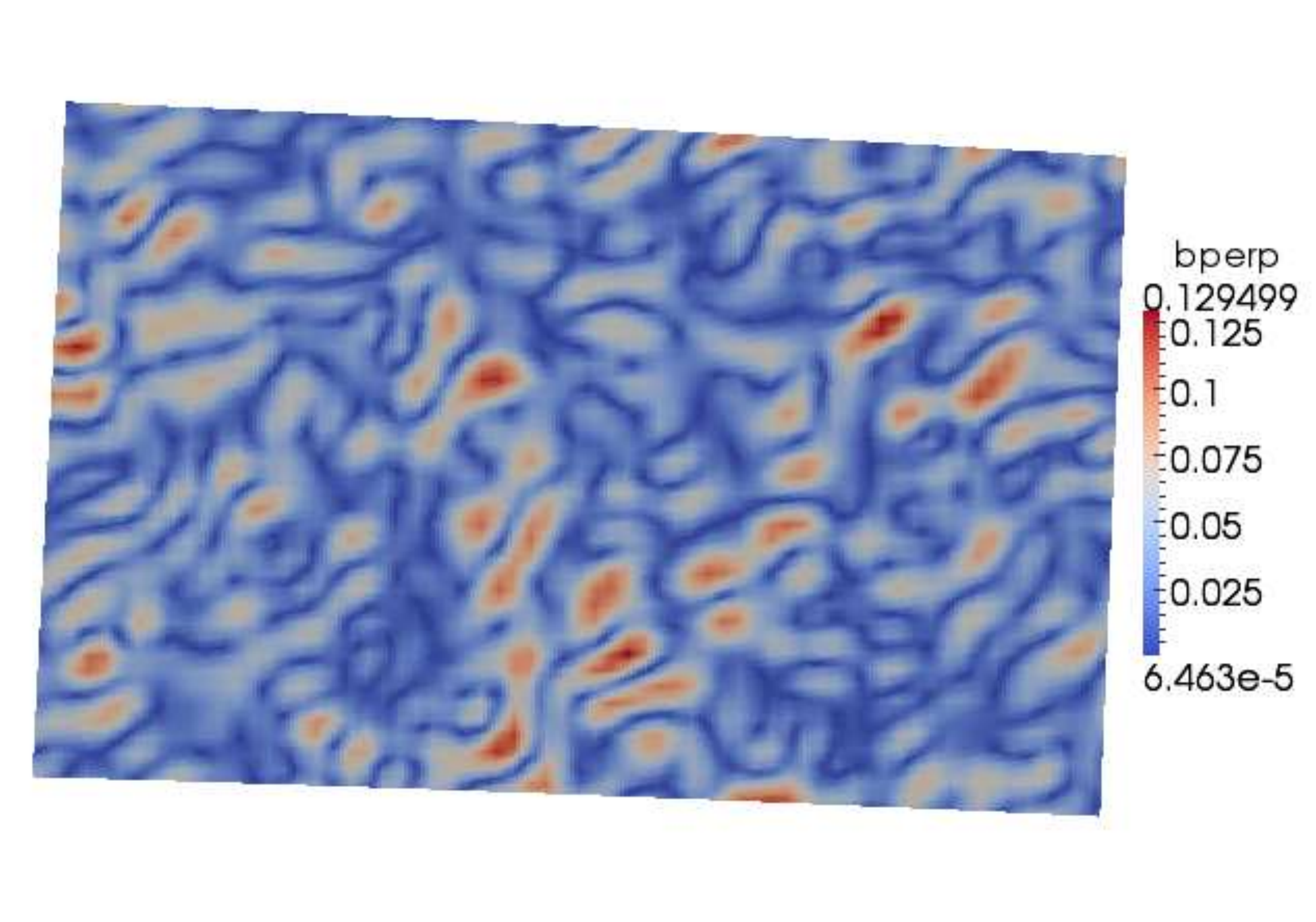}\includegraphics[trim=30 80 0 80,clip=true,width=0.45\textwidth]{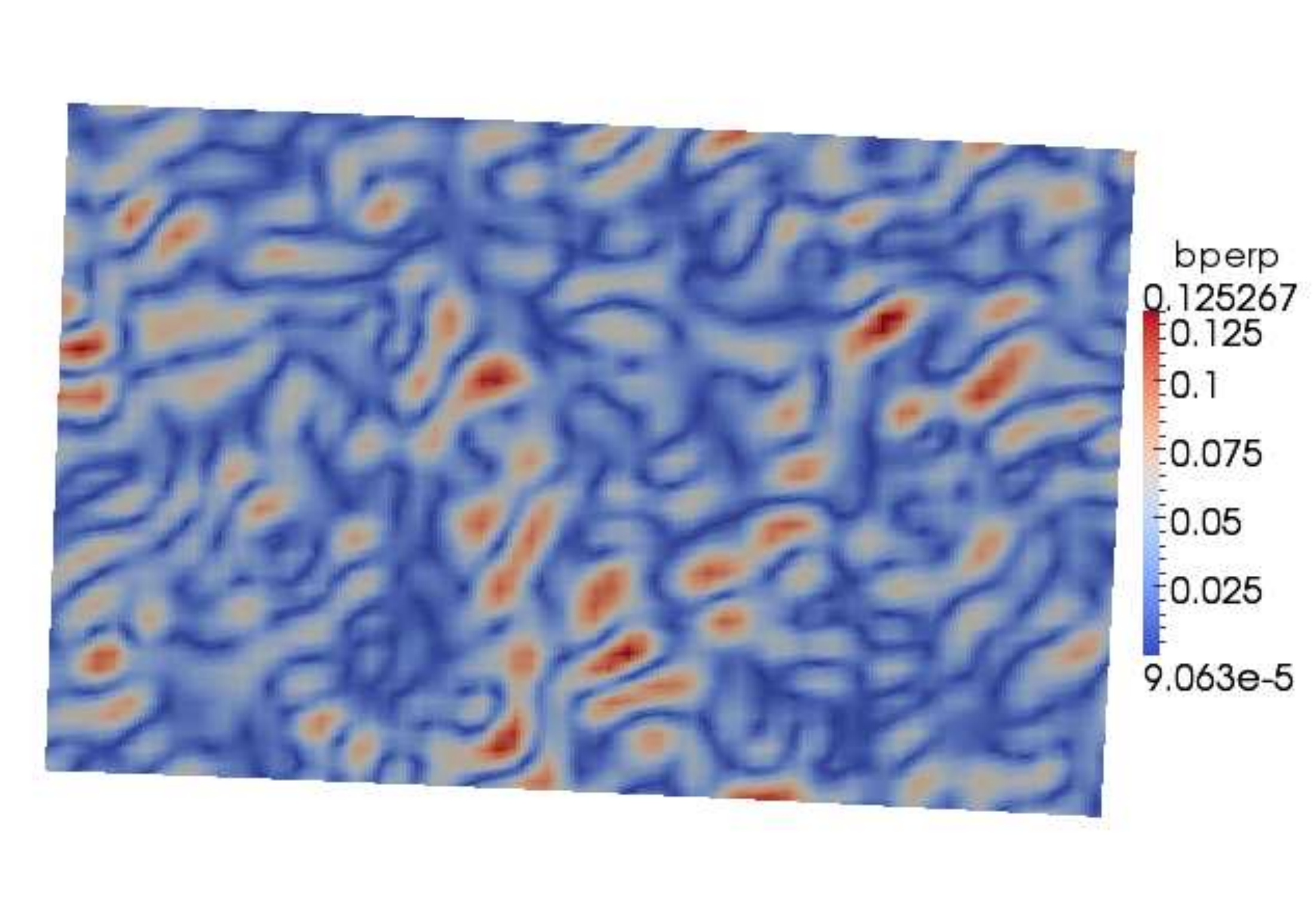}
 \includegraphics[trim=30 80 0 80,clip=true,width=0.45\textwidth]{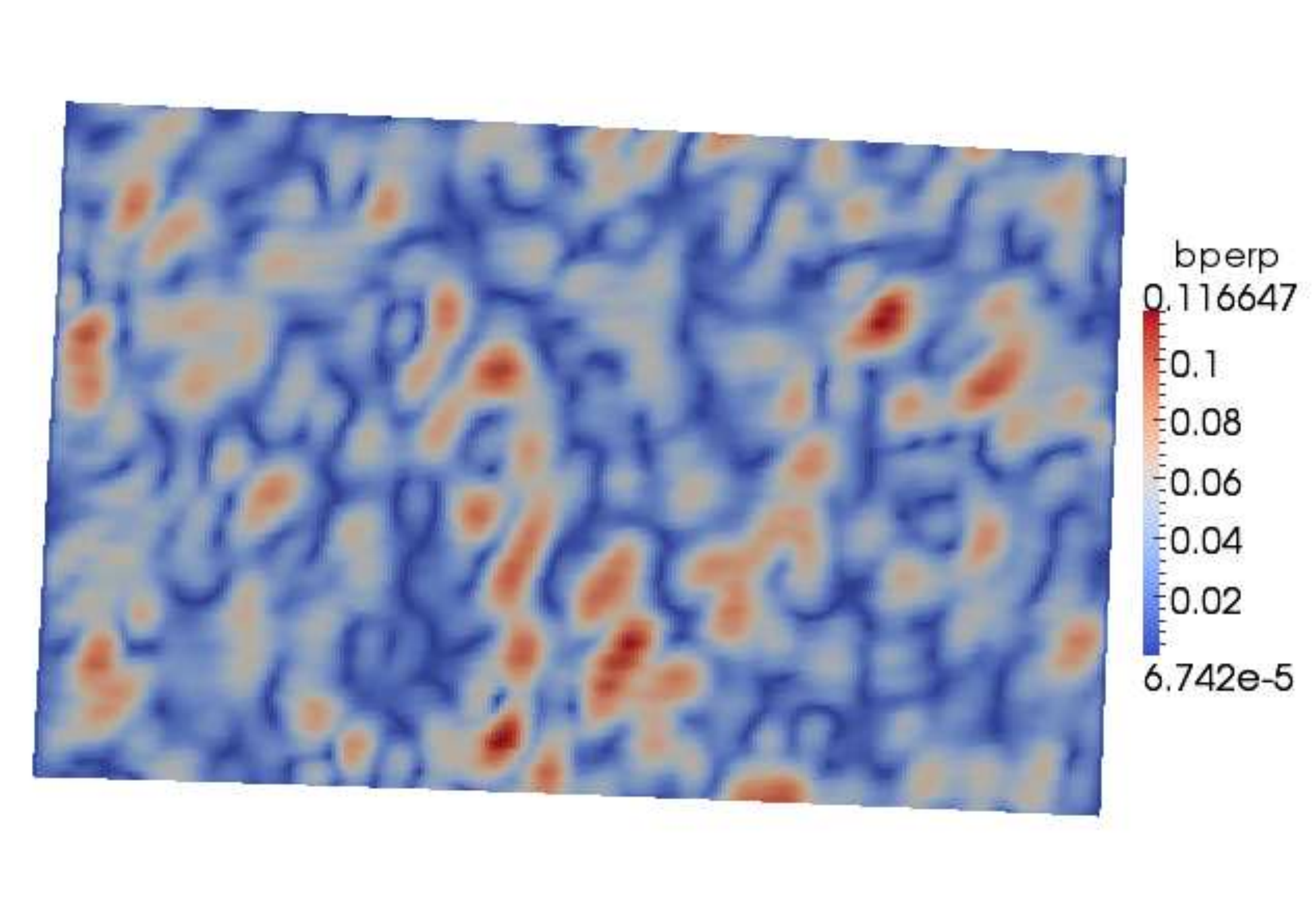}\includegraphics[trim=30 80 0 80,clip=true,width=0.45\textwidth]{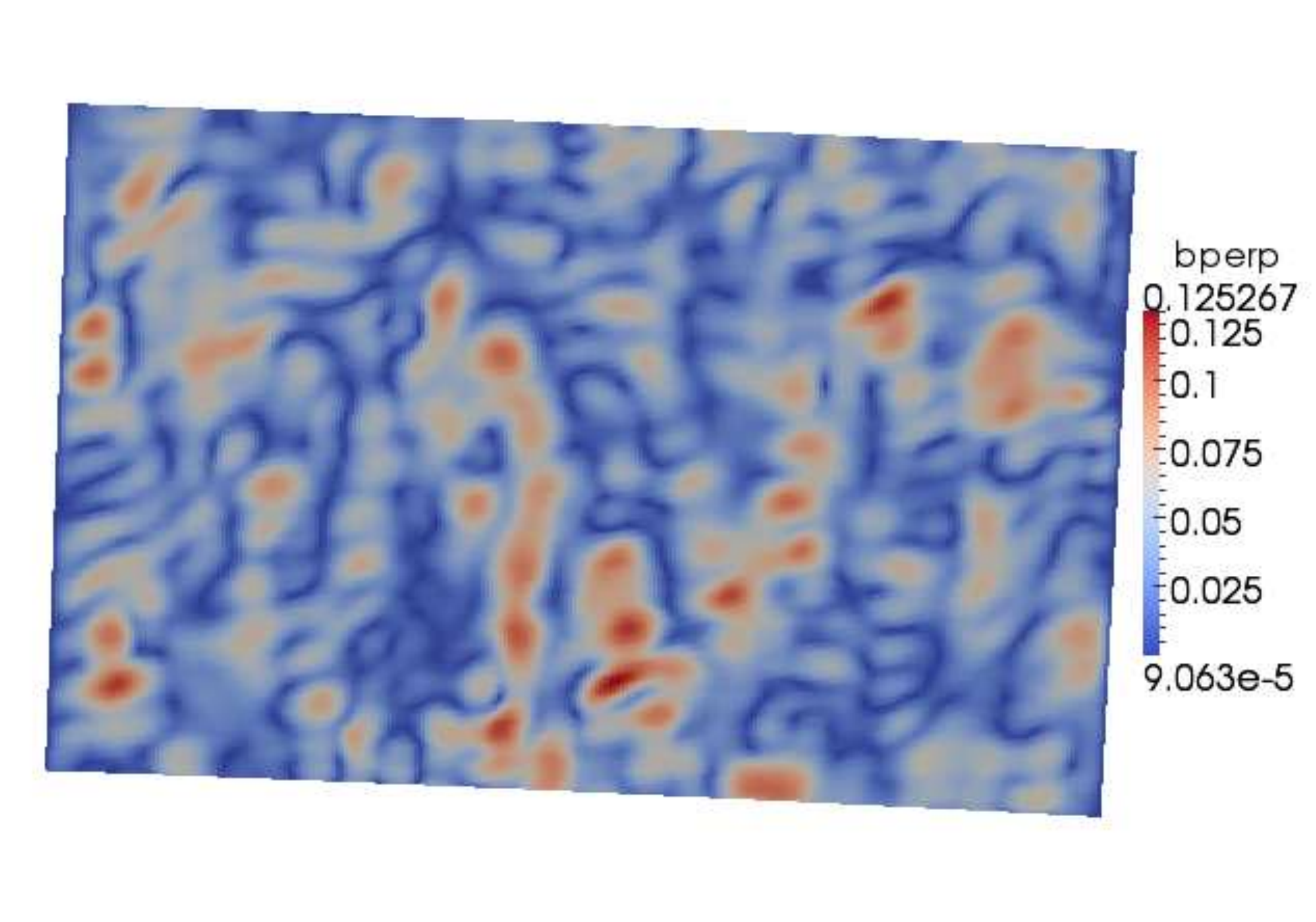}
 \includegraphics[trim=30 80 0 80,clip=true,width=0.45\textwidth]{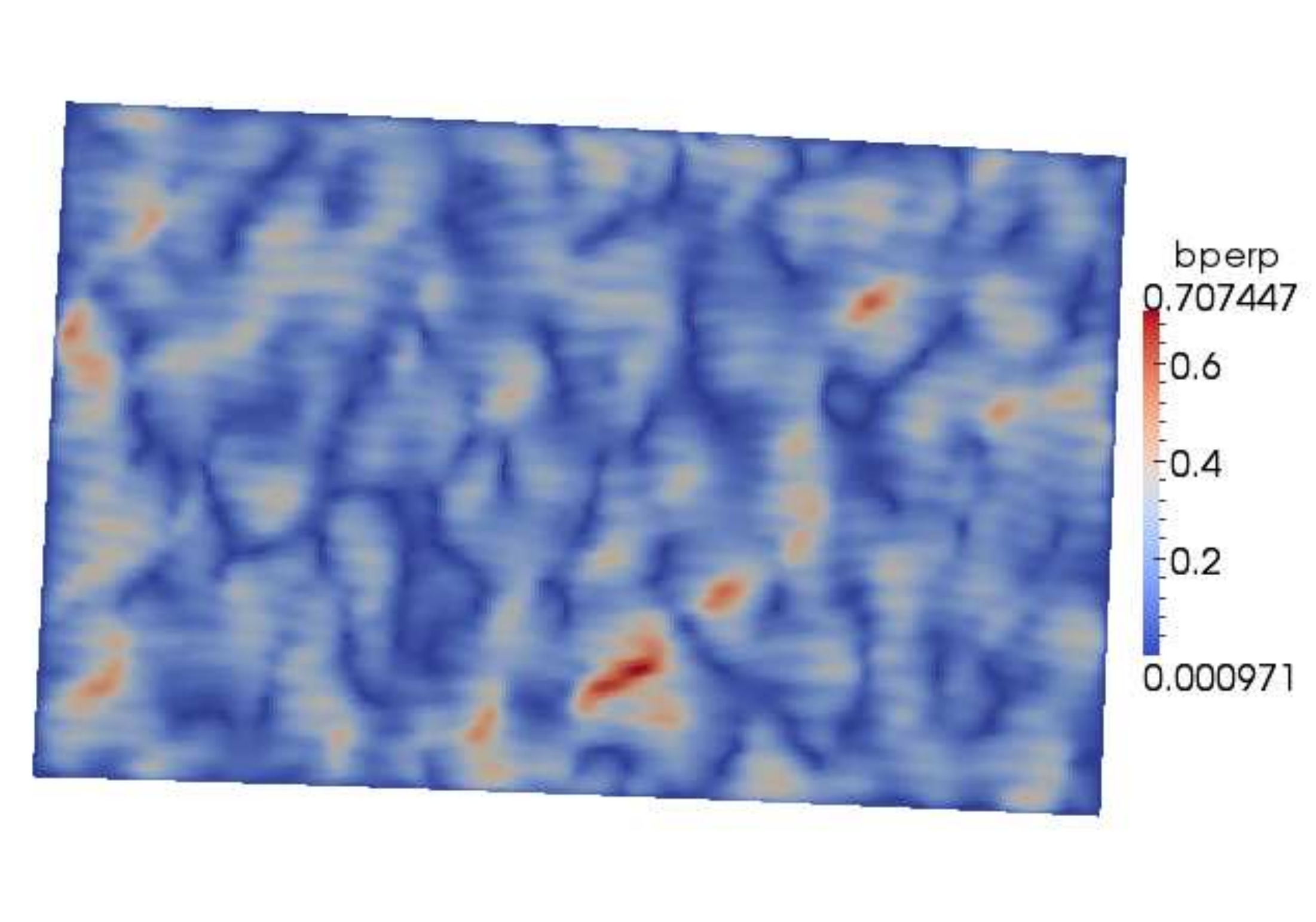}\includegraphics[trim=30 80 0 80,clip=true,width=0.45\textwidth]{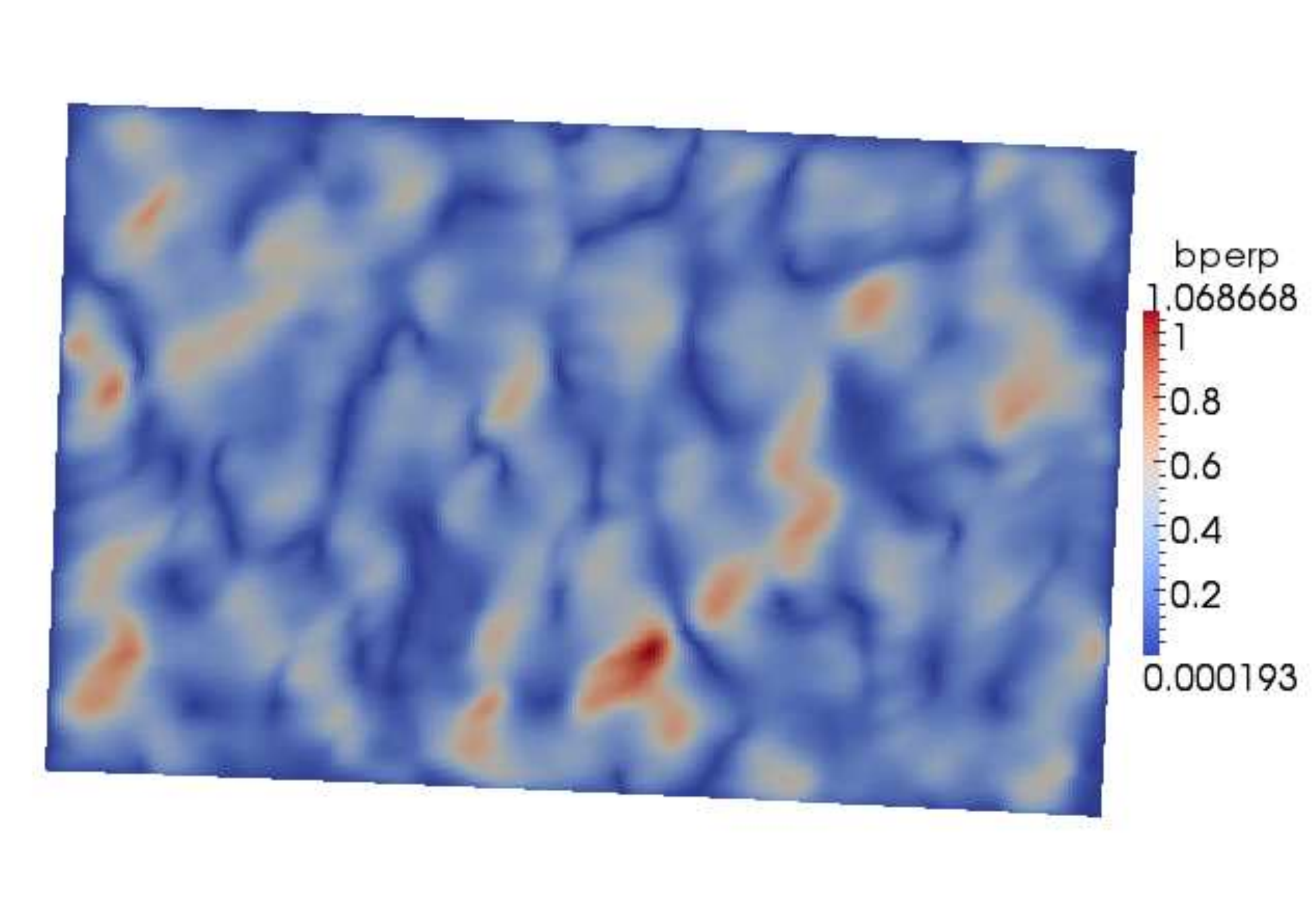}
 \includegraphics[trim=30 80 0 80,clip=true,width=0.45\textwidth]{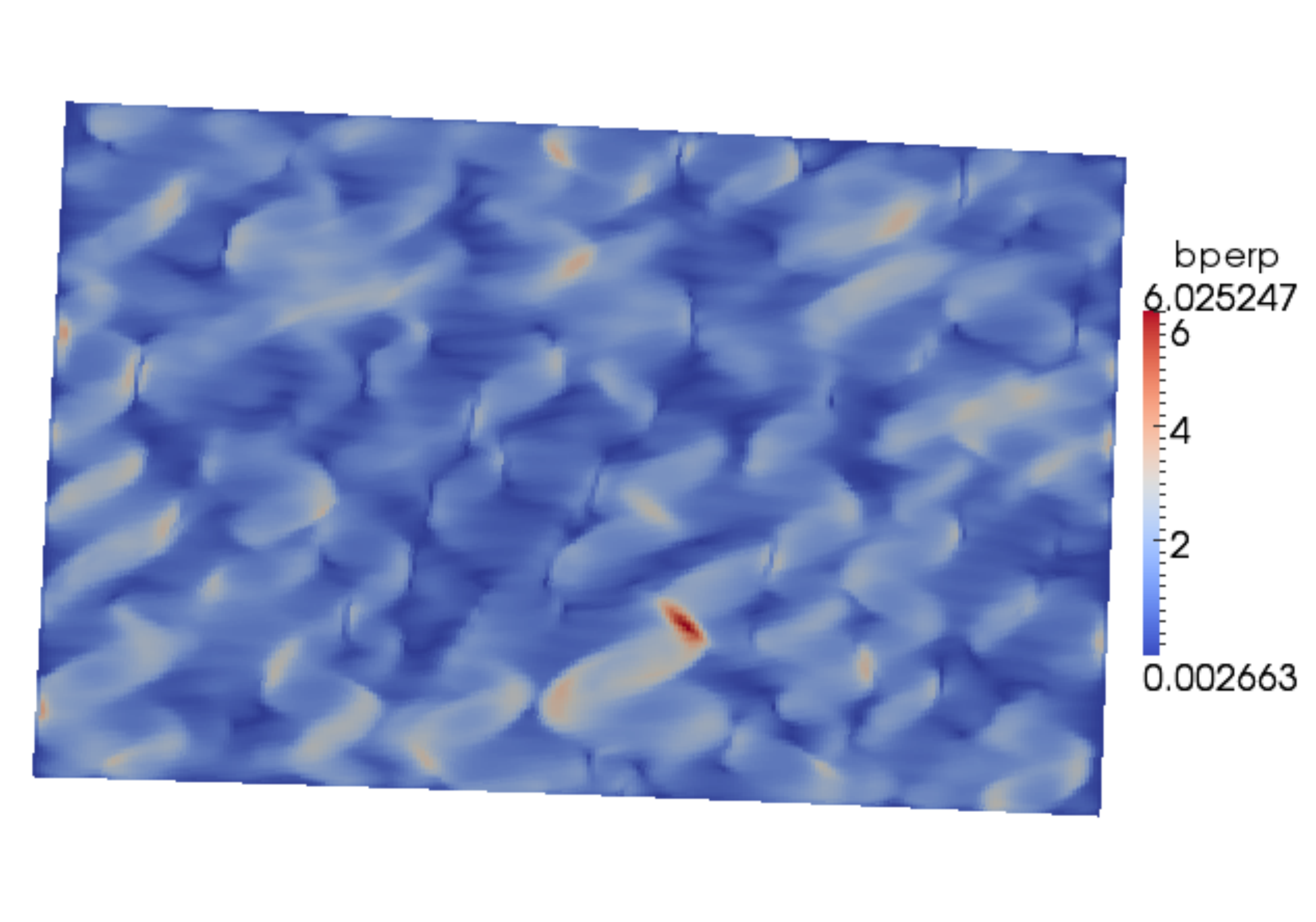}\includegraphics[trim=30 80 0 80,clip=true,width=0.45\textwidth]{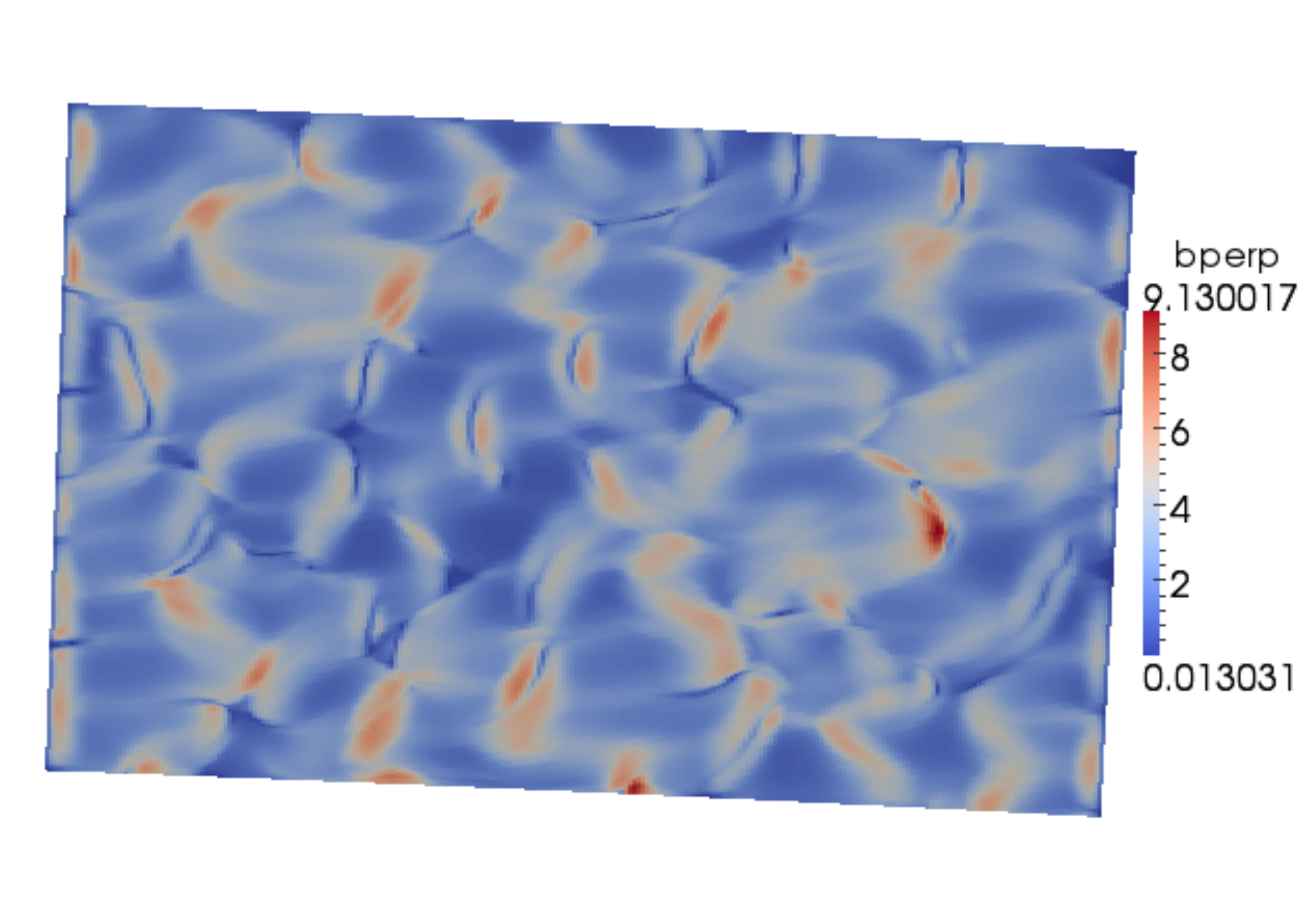}
 \includegraphics[trim=30 80 0 80,clip=true,width=0.45\textwidth]{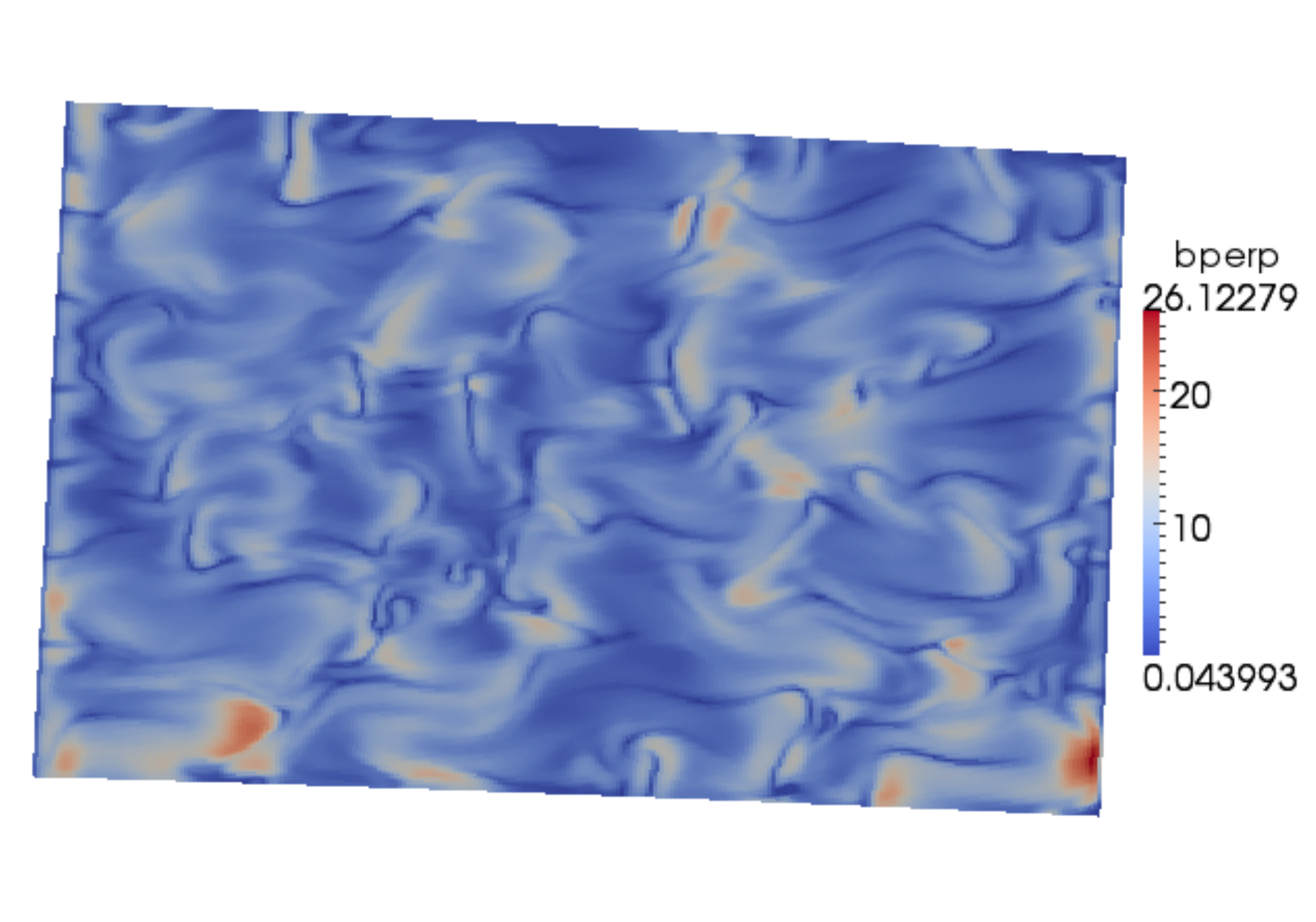}\includegraphics[trim=30 80 0 80,clip=true,width=0.45\textwidth]{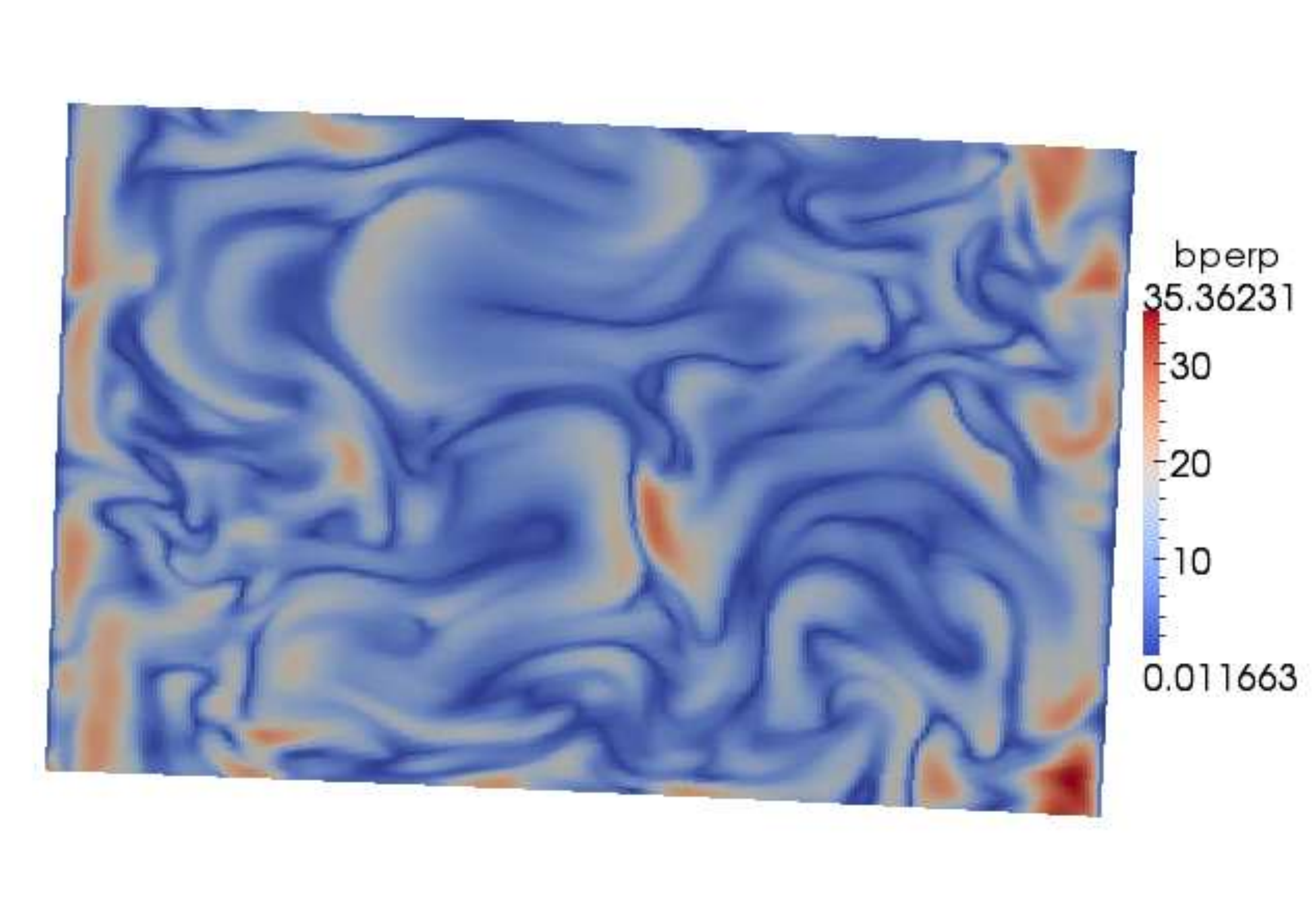}
 \caption[ ] {Evolution of $B_\perp$ as a function of time (top to bottom: 0, 40, 80, 120, 160 yrs) for different magnetic field strengths: the left column has $B_0=5\mu$G, the right column has $B_0=10\mu$G.
    \label{fig:smallscale}}
  \end{figure*}
  
Simulations on smaller scales give larger growth rates, as the growth rate goes up with wave number. In Figure~\ref{fig:smallscale} we present the evolution of the field strength for a simulation of just a small part of the upstream (for radius $r=[4.5-4.55]\times 10^{18}$cm). To further aid the instability we have increased the current to represent escape of CRs beyond $10^{12}$~eV. The left column shows snapshots of the simulation for a background magnetic field strength of $5~\mu$G, versus a magnetic field strength in the right column of $10~\mu$G. For both cases $\delta B/B_0$ is on the $5$\% level, with maximum fluctuations up to $15$\%. 

Growth of the instability is faster for the stronger magnetic field case, as the $j_0 \times B_\perp$ has a larger magnitude. Because of the lower magnetic field strength in the left column, the wave number of the fastest growing mode is significantly smaller than the scale of the initial perturbations, as can be seen especially in the third row. 
As the shock comes closer, the current increases. Together with the already amplified $B_\perp$ the growth proceeds rapidly to larger scales and amplitude. In the right column, values of $\delta B/B_0$ as high as $\sim35$ are reached. At this level, the self-consistency of the simulation can obviously be questioned. As the amplitude of fluctuations increase, in reality the cosmic rays will be confined closer to the shock, thereby decreasing the current of escaping cosmic rays. Additionally, the gyro radius may be smaller than the scale of the fluctuations, such that the assumption of a radial zeroth order current is not strictly valid. It remains to be investigated what happens when this feedback is included in simulations. 

\section{Discussion}
When the turbulent magnetic field is amplified, it confines cosmic rays more effectively to the shock region, thus allowing more efficient acceleration. The magnetic fluctuations grow most rapidly on the small scales, and increase in scale as the non-resonant hybrid instability progresses. The length scale of the field fluctuations determines the energy to which cosmic rays are confined. When cosmic rays are better confined, the current will decrease, thereby in turn lowering the growth rate of the instability.

\bibliography{../adssample}
\acknowledgements The research leading to these results has received funding
from the European Research Council under the European
Community's Seventh Framework Programme (FP7/2007-
2013)/ERC grant agreement no. 247039 and from grant number ST/H001948/1
made by the UK Science Technology and Facilities Council. 
\end{document}